\documentclass[a4,11pt,aps,floats,showpacs]{article}
\usepackage{graphicx}
\usepackage{pst-coil}
\usepackage{amsmath}
\usepackage{verbatim}

\def\be{\begin{equation}}
\def\ee{\end{equation}}
\def\bea{\begin{eqnarray}}
\def\eea{\end{eqnarray}}
{
{
\def\mpl{M_{\rm {Pl}}}
\def\Mpc{{\rm Mpc}}
\def\gev{{\rm \,Ge\kern-0.125em V}}
\def\tev{{\rm \,Te\kern-0.125em V}}
\def\mev{{\rm \,Me\kern-0.125em V}}
\def\ev{\,{\rm eV}}

\def\T1{T_1}

\def\half{\frac{1}{2}}

\begin{document}
\title{{{\bf WMAPping the Inflationary Universe}}
\author{Raghavan Rangarajan \\ 
Physical Research Laboratory\\ Ahmedabad 380 009, India\\
raghavan@prl.res.in}\\
}
\date{}
\maketitle
\begin{abstract}
An epoch of accelerated expansion, or inflation, in the early universe solves
several cosmological problems.  While there are many models of inflation 
only recently has it become possible to discriminate between some of the models
using observations of the cosmic microwave background radiation 
and large-scale structure.  In this talk, we discuss inflation and its
observational consequences,
and then the status of current cosmological observations and 
their implications for different 
models of
inflation.
\end{abstract}

\section{Introduction}

Inflation is a period of rapid expansion in the early universe that
resolves several cosmological problems.  While 
many different models of inflation such as new inflation, chaotic inflation, natural inflation and
hybrid inflation have been proposed over the past two and a half decades,
we are now entering the era where some
models can be ruled out or shown to be consistent with observations.  
In this
talk we shall first discuss inflation and its 
observational consequences. 
We shall then discuss
what the current observations of the 
cosmic microwave background radiation (CMBR) and large-scale structure
(LSS) in the universe
imply for models of inflation.

\section{Inflation and its observational consequences}

Formally, inflation can be divided into two eras - the inflationary era
and the reheating era.  In the inflationary era the universe undergoes
accelerated expansion because of the dominant potential energy of a slowly
moving scalar field $\phi$ (the inflaton).  In the subsequent reheating
era, the inflaton decays and reheats the universe.  Inflation provides an
explanation as to why the CMBR is (nearly) isotropic (the horizon
problem) and why the universe today is spatially flat (the flatness problem).
Furthermore it dilutes away unwanted relics such as GUT monopoles.

In addition to the above effects,
quantum fluctuations of the inflaton during inflation
give rise to fluctuations in the energy density of particle species after the
inflaton decays.  Furthermore, quantum (tensor) 
fluctuations of the metric during inflation give rise to a cosmic
gravitational wave background.

Fluctuations in the density of non-relativistic particles, or matter, are
the seed for large-scale structure observed in the universe today.
If $\delta_k$ is the Fourier transform of $\delta \rho_m/\rho_m$ then the
matter power spectrum is given as $P(k)=|\delta_k|^2$.  
Inflation predicts that $P(k)=A k^{n}$ with the spectral index $n\sim1$.
[If $n=1$, $P(k)$ is called a scale-invariant or Harrison-Zeldovich spectrum.]
A related quantity is the scalar power spectrum, $P_S(k)=A_S k^{n-1}$,
which is associated with the scalar curvature.
More generally, inflation gives
\begin{equation}
n(k)=n(k_0) + \half dn/d(\ln k_0) \ln(k/k_0)\, ,
\end{equation}
which is referred to as running of the spectral index.
Different models of inflation give slightly different behaviour for $n(k)$.
Simulations of large-scale structure for an inflationary
universe in the context of any particular inflation model
can be compared
with large-scale structure data from the Sloan Digital Sky Survey (SDSS),
2 Degree Field Galaxy Redshift Survey (2dFGRS), etc.

Fluctuations in the photon density and interactions of the photons with the
fluctuations in the matter at decoupling give the anisotropies in the CMBR 
temperature (first detected by the Cosmic Background Explorer (COBE) in 1992).
The 2-point correlation function for photons is
\begin{equation}
\frac{\langle T(\hat n_1)T(\hat n_2)\rangle_\theta - T_0^2}{T_0^2},
\end{equation} 
where $T(\hat n)$ is the temperature in direction $\hat n$, $T_0$ is the 
mean temperature and $\langle .. \rangle_\theta$ indicates averaging over
all points in the sky separated by an angle $\theta$. 
The shape of the 2-point function depends on $P(k)$ which is provided by
inflation.
The matter power spectrum in inflation models imply 
a flat 2-point temperature correlation function at
large angles and peaks at smaller angles.

In addition to the above form of the 2-point correlation function, inflation 
predicts that the fluctuations in the CMBR will be largely Gaussian.  The
$n$-point temperature correlation function
$\langle T(\hat n_1) T(\hat n_2) ... T(\hat n_n) \rangle
\propto \langle \phi(x_1) \phi(x_2) ... \phi(x_n)\rangle$.
The small value of the 2-point temperature correlation function 
($\sim O(10^{-5}$) at large angles) implies that
the inflaton is a very weakly coupled field \cite{kt}.  
Hence the $n$-point correlation
in $\phi$ is approximately 0 for odd $n$ and
$
\sim \langle \phi(x_1) \phi(x_2)\rangle ...\langle\phi(x_{n-1}) \phi(x_n)\rangle
$
for even $n$.
Thus higher point correlations in the temperature are 0 or powers of the 
2-point correlation function, indicating Gaussian fluctuations.

Tensor fluctuations of the metric generated during inflation are associated
with the gravitational wave background.  The tensor power spectrum is given
by
$P_T(k)=A_T k^{n_T}$.
The ratio of the tensor and scalar power spectra 
is denoted as $r$.  Now the
scale of inflation is equal to
$(r/0.07)^{1/4}\,1.8\times 10^{16}\gev$,
and so a detection of gravitational waves can give the scale of inflation.

The gravitational wave background affects the CMBR and contributes to
the 2-point correlation function.
Furthermore, 
the interaction of gravitational waves with photons gives a quadrupole
anisotropy in the photon distribution which leads to polarisation in the
CMBR after Thomson scattering off electrons at decoupling.

The current observations of the CMBR and LSS 
broadly imply the following in relation
to inflation:

\begin{itemize}
\item The matter density fluctuations are nearly scale-invariant with
$P(k) = A k^{n}$ with $n\sim 1$

\item The presence of peaks in the CMBR fluctuations are consistent with 
inflation and not with other mechanisms of producing primordial fluctuations
such as cosmic strings, etc.

\item There is no evidence for non-Gaussianity in the temperature 
fluctuations \cite{creminellietal}

\item There are correlations on super-horizon scales, i.e., on scales larger
than the horizon at decoupling, in the polarisation spectrum indicating the 
presence of fluctuations on scales larger than the horizon size at decoupling.
The presence of such seemingly acausal fluctuations is a prediction of 
inflation.  (Correlations on super-horizon scales are also seen in the 
temperature spectrum.  However these can be generated causally after
decoupling from sub-horizon
scale fluctautions via the integrated Sachs-Wolfe effect.)

\end{itemize}

Thus the inflationary paradigm is consistent with current observations.
With regards to gravitational waves generated during inflation, 
there has been no direct detection yet.
Furthermore, there
are other sources of the quadrupole anisotropy at decoupling 
and the polarisation
detected so far does not provide the value of the
gravitational wave contribution.
We now look at specific values of cosmological parameters inferred from
the CMBR and LSS data to discriminate between different models of inflation.

\section{Distinguishing between different inflation models}

Inflation models can be parametrised by the following slow roll parameters,

\begin{equation}
\epsilon_V= \frac{\mpl^2}{16\pi}\left(
\frac{V^\prime(\phi)}
{V(\phi)}
\right)^2
\,\,\,\,\,
\eta_V= \frac{\mpl^2}{8\pi}\left(\frac{V^{\prime\prime}(\phi)}
{V(\phi)}
\right)
\,\,\,\,\,
\xi_V= \left(\frac{\mpl^2}{8\pi}\right)^2 \left(\frac
{V^\prime V^{\prime\prime\prime}}
{V^2}
\right)
\end{equation}
and higher derivatives.  Above $V(\phi)$ is the inflaton potential and
$\prime$ refers to a derivative with respect to $\phi$.
Note that different authors use different definitions of the slow roll
parameters; ours are taken from Ref. \cite{peirisetal}.
The assumption that the inflaton rolls slowly during the inflationary
epoch is equivalent to $\epsilon_V,\eta_V,\xi_V\ll1$.  Now,
$n  =  1-6\epsilon_V+2\eta_V,\,
r =  16\epsilon_V$ and
${dn}/{d(\ln k)}  =  16\epsilon_V\eta_V-24\epsilon_V^2-2\xi_V$.
Therefore
constraints on $n,{dn}/{d(\ln\,k)}$ and $r$ from CMBR and LSS data
give
limits on $\epsilon_V,\eta_V$ and $\xi_V$.
Below we first present constraints on $n,{dn}/{d(\ln\,k)}$ and $r$.

WMAP3 alone is consistent with \cite{kkmr2006}
\begin{equation}
0.94 < n < 1.04 \,\,\,\,\,dn/d(\ln k) = 0\,\,\,\,\, {\rm and} \,\,\,\, r<0.60
\end{equation}
For a running spectral index \cite{kkmr2006}
\begin{equation}
1.02 < n < 1.38 \,\,\,\,\, -0.17 < dn/d(\ln k) < -0.02 \,\,\,\,\,{\rm and}
\,\,\,\,\,
r<1.09
\end{equation}
The parameters 
are estimated at a pivot scale $k_*=0.002 \Mpc^{-1}$ in 
Ref. \cite{kkmr2006}.
Note that $dn/d(\ln k) < -0.02$ rules out all single field slow roll models
as it leads to an insufficient duration of inflation \cite{eastherpeiris}.

Combining results from WMAP3 and SDSS gives \cite{kkmr2006}
\begin{equation}
0.93 < n < 1.01 \,\,\,\,\,dn/d(\ln k) = 0\,\,\,\,\, {\rm and} \,\,\,\, r<0.31
\end{equation}
or
\begin{equation}
0.97 < n < 1.21 \,\,\,\,\, -0.13 < dn/d(\ln k) < 0.007 \,\,\,\,\,{\rm and}
\,\,\,\,\,r<0.38
\end{equation}
Note that the upper limit on $r$ has decreased.  This is because as $r$
increases, the scalar amplitude decreases and for $r\ge0.3$ this then
adversely affects the LSS.

Including small angle CMBR data from CBI, ACBAR, VSA and B2K with WMAP3 and
2dFGRS LSS data gives \cite{frm2006}
\begin{equation}
0.95 < n < 0.98 \,\,\,\,\,dn/d(\ln k) = 0
\end{equation}
or
\begin{eqnarray}
0.94 < n < 1.09 \,\,\,\,\,& -0.14 < dn/d(\ln k) < -0.013 
\end{eqnarray}
The corresponding upper limit on $r$ of 0.26
gives an upper limit on the scale of inflation of
$2\times10^{16}\gev$.  In Ref. \cite{frm2006}
the pivot scale $k_*=0.01 \Mpc^{-1}$.

\vspace{0.1in}
{\bf {\large Implications for inflation models}}
\vspace{0.1in}

Different models of inflation are distinguished by the form of
the inflaton potential and therefore correspond to different ranges of
values for the slow roll parameters \cite{kkmr2006}.  Thus the above
constraints on spectral parameters provide constraints on inflation
models.

{\bf New inflation:} In new inflation the inflaton potential has the form
\begin{equation}
V=V_0\left[1-\left(\frac{\phi}{\mu}\right)^p\right]
\end{equation}
with the initial
position of the inflaton at small values of $\phi$.  An example of such a
potential is the Coleman-Weinberg potential which is approximately given by
$V=V_0-\lambda\phi^4$
in the small $\phi$ region.
New inflation models with $p\ge3$ are consistent with current observations
\cite{alabidilyth2006}.

{\bf Chaotic inflation:}  Chaotic inflation models have a potential of the
form $\sim A\phi^p$ 
with the inflaton field initially
displaced far from the minimum of its potential.  The initial value of the field
is greater than $\mpl$ (though $V(\phi)<\mpl^4$).  
The field rolls slowly till
$\phi\sim 0.1\,\mpl$ during the inflationary era, 
and then oscillates in its potential and decays.
Models with $V(\phi)\sim m^2\phi^2$ are consistent with the data while models
with $V(\phi)\sim\lambda\phi^4$ are ruled out by WMAP3 and LSS 
data \cite{kkmr2006}.
However, recently it has been pointed out that chaotic inflation models with
$V(\phi)\sim\lambda\phi^4$ are still allowed if the neutrino fraction
$f_\nu\equiv \Omega_\nu/\Omega_c=0.03-0.05$.  This further implies that 
$\sum_i m_{\nu i} = 0.3-0.5 \ev$,
as in quasi-degenerate nuetrino mass models \cite{hamannetal}.  
This is an interesting consistency
relation between a model of inflation and a model of neutrino masses.

{\bf Natural inflation:} In natural inflation models the inflaton is a
pseudo-Nambu-Goldstone boson
associated with spontaneous symmetry breaking at a scale $f$
and small explicit (dynamical) symmetry breaking of order $\Lambda$.
In these models the flat potential of a Nambu-Goldstone boson $\phi$ is tilted
because of explicit symmetry breaking and
\begin{equation}
V(\phi)=\Lambda^4[1+\cos(\phi/f)] \,.
\end{equation}
The tilt of the potential $\sim$ height/width $\sim (\Lambda/f)^4$
and for $\Lambda\sim 10^{15}\gev$ and $f\sim 10^{19}\gev$ the potential is
flat enough to satisfy constraints on the inflaton potential.  
Because the flatness of the 
potential is associated with a naturally flat Nambu-Goldstone boson potential
and small dynamical symmetry breaking, 
rather than an unnaturally small coupling, this scenario is called natural 
inflation.  Natural inflation is consistent with the 
WMAP3 data \cite{savageetal}.

{\bf Hybrid inflation:}  These models involve a potential with 2 fields
$\phi$ and $\chi$.  The potential has the form
\begin{equation}
V(\phi,\chi)=\lambda (\chi^2-\chi_0^2)+\half m^2\phi^2 +\half\lambda^\prime
\chi^2\phi^2 \,.
\end{equation}
$\phi$ is initially displaced from the minimum of its potential and rolls
slowly in its potential.
For $\phi>\phi_1$,
$\chi$ is localised near the 
origin
and 
inflation is driven by the energy density of $\phi$ or the false vacuum energy
of $\chi$ (in the latter stages).
When $\phi$ crosses the threshhold value $\phi_1$ the potential for $\chi$ 
turns over and now $\chi$ 
rolls down towards the new minimum $\chi_0$.  
This ends the inflationary era and
$\phi$ and $\chi$ then 
oscillate in their potential and decay. 
The original non-SUSY version of this model 
implies that $n\ge1$ \cite{lindehybrid} and is ruled
out if indeed $n<1$.
However hybrid inflation models in the context of SUSY and SUGRA, referred to
as D-term and F-term inflation, can give $n<1$ and there does exist a region of
parameter
space consistent with the data \cite{jeannerot}.

\section{Conclusion}

In summary, the inflationary paradigm is consistent with the CMBR and LSS data.
However, while
current data on CMBR and LSS provide information on the scalar and tensor
power spectra making it possible to discriminate between some models of 
inflation,
it is still not possible to rule out many specific
models of inflation.
In the future,
the European Space Agency's Planck mission will give
better data on the spectral index and its running which will help in
constraining inflation models.  For example, a detection of
large negative running
of the spectral index will rule out single field slow roll inflation models.
Furthermore,
Planck and the ground-based Clover experiment will measure $r$ upto $10^{-2}$.
This will help to constrain the scale of inflation.


\end{document}